\documentclass[pra,twocolumn,showpacs,preprintnumbers,amsmath,linenumbers,amssymb]{revtex4}
\usepackage{amsfonts}
\usepackage{mathrsfs}
\usepackage{amsmath}
\usepackage{amssymb}
\usepackage{txfonts}

\usepackage{graphicx}
\usepackage{dcolumn}
\usepackage{bm}
\usepackage[dvipdfm]{xcolor}
\usepackage{subfigure}
\usepackage{textcomp}

\begin{document}



\title{Active optical clock based on four-level quantum system}

\author{Tonggang Zhang, Yanfei Wang, Xiaorun Zang, Wei Zhuang, Jingbiao Chen}\thanks{E-mail: jbchen@pku.edu.cn}
\affiliation{Institute of Quantum Electronics and State Key
Laboratory of Advanced Optical Communication Systems $\&$ Networks,
School of Electronics Engineering $\&$ Computer Science, Peking
University, Beijing 100871, China}


\begin{abstract}
Active optical clock, a new conception of atomic clock, has been
proposed recently. In this report, we propose a scheme of active
optical clock based on four-level quantum system. The final accuracy
and stability of two-level quantum system are limited by
second-order Doppler shift of thermal atomic beam. To three-level
quantum system, they are mainly limited by light shift of pumping
laser field. These limitations can be avoided effectively by
applying the scheme proposed here. Rubidium atom four-level quantum
system, as a typical example, is discussed in this paper. The
population inversion between $6S_{1/2}$ and $5P_{3/2}$ states can be
built up at a time scale of $10^{-6}$s. With the mechanism of active
optical clock, in which the cavity mode linewidth is much wider than
that of the laser gain profile, it can output a laser with
quantum-limited linewidth narrower than 1 Hz in theory. An
experimental configuration is designed to realize this active
optical clock.
\end{abstract}

\pacs{32.80.Xx, 06.30.Ft, 42.55.-f, 37.30.+i}

\maketitle

\section{INTRODUCTION}

Optical clocks have demonstrated great
improvements in stability and accuracy over the microwave frequency standards.
Since the proposal of active optical clock~\cite{Chen1,Chen2, Wang},
a number of neutral atoms with two-level, three-level and four-level at thermal beam,
laser cooling and trapping configurations have been investigated
recently~\cite{Chen1, Chen2, Wang, Zhuang1, Zhuang2, Zhuang3, Yu, Chen3, Meiser1, Meiser2, Xie, Zhuang4, Zhuang5}.
The potential quantum-limited linewidth of active optical clock is narrower
than mHz, and it is possible to reach this unprecedented linewidth since the thermal
noise of cavity mode can be reduced dramatically with the mechanism of
active optical clock. It has been recognized that active optical clock has the potential to improve
the stability of the best atomic clocks by about 2 orders of magnitude~\cite{Meiser1, Meiser2, Uwe}.

Until now, the major experimental schemes of active optical clock are based on trapped quantum system and
atomic beam quantum system. To the latter, the residual Doppler shift will be the main limitation, thus the final accuracy and
stability of two-level quantum system are limited by second-order Doppler shift of thermal atomic beam. Laser cooled and trapped
quantum system provides a solution to this limitation.
As for three-level quantum system, the system stability and accuracy  are mainly affected by the light shift of pumping laser.
Four-level quantum system can overcome these limitations and thus
will be a better choice for active optical clock with improved performance~\cite{Zhuang3}.

We have investigated several alkali metals including potassium, rubidium and cesium
and found that the four-level quantum systems of these elements are appropriate choices based on
the mechanism of active optical clock. Here we take rubidium atom as an example.
The population inversion can be realized as shown in Fig.~1.

\begin{figure}
\label{fig:Figer1}
\includegraphics[width= 8 cm]{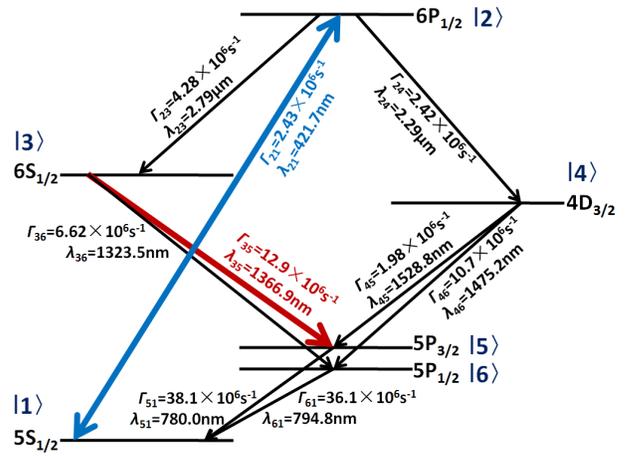}
\caption{(Color online) Scheme of Rb four-level quantum system active optical clock
operating at 1366.9 nm clock transition,
pumped with 421.7 nm laser~\cite{NIST}.}
\end{figure}

A 421.7 nm laser operating at $5S_{1/2}$ to $6P_{1/2}$ transition is used as pumping laser.
First, the atoms at $5S_{1/2}$ state are pumped to
$6P_{1/2}$ state, and then decay to $5S_{1/2}$, $6S_{1/2}$ and  $4D_{3/2}$ states by spontaneous transitions.
Second, the atoms at $6S_{1/2}$ and $4D_{3/2}$ states decay to $5P_{1/2}$ and $5P_{3/2}$ states respectively.
Third, the atoms at $5P_{1/2}$ and $5P_{3/2}$ states return to ground state $5S_{1/2}$ and then are pumped to
$6P_{1/2}$ state by pumping laser again. This process continues until the population
inversion is established between $6S_{1/2}$ and $5P_{3/2}$ states since the lifetime of 6S state is longer
than that of 5P state.

One should note that the population inversions are also built up
between $6S_{1/2}-5P_{1/2}$, $4D_{3/2}-5P_{3/2}$ and
$4D_{3/2}-5P_{1/2}$ states. We choose $6S_{1/2}$ and $5P_{3/2}$
states as the object of study in this paper because the population
inversion between them is the greatest according to our calculation.
At the same time, one should pay attention to the fact that the
split of the two hyperfine levels of $5P_{1/2}$ state is wider than
that of $5P_{3/2}$ state. So if we want to realize this active
optical clock in experiment, the transition between $6S_{1/2}$ and
$5P_{1/2}$ states will be a better selection for avoiding modes
competition.

Once the population inversion is built up, a Fabry-Perot type
resonator, with a bad cavity whose cavity mode linewidth is much
wider than that of the laser gain profile, is used to realize active
optical clock laser output. Using the definition of
$a=\Gamma_{c}/\Gamma_{gain}$~\cite{Chen1,Chen2, Kuppens}, here
$\Gamma_{c}$ is cavity loss rate and $\Gamma_{gain}$ is gain
medium linewidth. According to the mechanism of active optical
clock~\cite{Chen1}, the Fabry-Perot resonator should be pushed deep
down into the bad-cavity regime $a\gg1$. The quantum-limited
linewidth of active optical clock based on four-level quantum system
is expected to be narrower than Hz.

In this report, we will present the result of the calculations of dynamical process of
population inversion, and lasing process with a bad cavity. An experimental
scheme is designed to realize Rb four-level quantum system active optical clock.

\section{POPULATION INVERSION OF RUBIDIUM ATOMS FOUR-LEVEL QUANTUM SYSTEM}

A pumping laser operating at 421.7 nm is used to realize population inversion of Rb four-level quantum system.
The Rabi frequency of pumping laser is $\Omega=d\cdot\varepsilon/\hbar$, where $d$ is electric dipole matrix
between $5S_{1/2}$ state and $6P_{1/2}$ state, $\varepsilon$ is electric field strength
of the pumping laser. The spontaneous decay rate is $\Gamma_{21}=\omega_{21}^{3}d^{2}/(3\pi\hbar c^{3}\varepsilon_{0})$
where $\omega_{21}=2\pi c/\lambda_{21}$. The relationship between
light intensity of pumping laser and the Rabi frequency is
$I=2\pi h c \Omega^{2}/(3\lambda_{21}^{3}\Gamma_{21})$.
If the pumping laser is 10 mW with beam waist 1 mm, and considering that
$\Gamma_{21}=2.43$ MHz, the pumping rate
from $5S_{1/2}$ to $6P_{1/2}$ is $\Omega=3.7\times10^{7}$ s$^{-1}$.
The saturation light intensity is
$I_{s}=\pi h c \Gamma_{21}/(3\lambda_{21}^{3})=0.7$ mW/cm$^{2}$,
and $\Omega_{s}=1.7\times10^{6}$ s$^{-1}$ for $I=I_{s}$.
So the Rabi frequency $\Omega$ can be set from $1.7\times10^{6}$ s$^{-1}$ to about $3.7\times10^{7}$ s$^{-1}$.

\begin{figure}[!h]
\label{fig:Figer2}
\includegraphics[width= 8 cm]{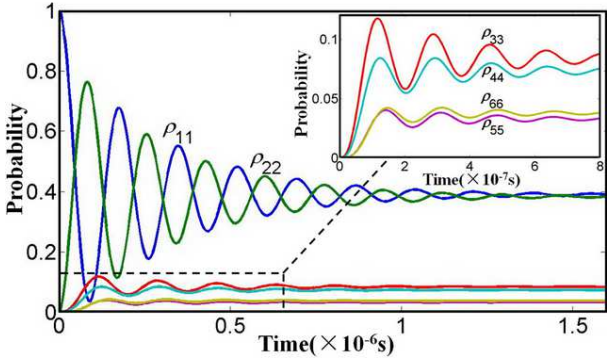}
\caption{(Color online) The dynamical populations of Rb atom with Rabi
frequency $\Omega=1.8\times10^{7}$ s$^{-1}$.}
\end{figure}

Supposing that atoms are within a thermal gas cell or
magneto-optical trap (MOT), part of the atoms would be pumped to
$6P_{1/2}$ state. Atoms at $6P_{1/2}$ state spontaneously decay to
metastable states $6S_{1/2}$ and $4D_{3/2}$ and continue decaying to
the lower sate. The atoms that return to $5S_{1/2}$ state are
repumped by the pumping laser. When the 421.7 nm pumping laser is
operating on one of the two hyperfine structure energy levels of 5S ground state, a
repumping laser can be added to close the atoms leakage to the other
hyperfine structure energy level. We will not discuss
the repumping laser function in this paper for simplicity.
Finally, the population inversion could be established between $6S_{1/2}$
and $5P_{3/2}$ states. The whole process is clear after an approximate
calculation in the following.

The density matrix equations for Rb atoms in a thermal gas cell
interaction with the pumping light, based on the theories of
interaction between atoms and light, are written in RWA-approximation as follows:
\begin{equation}
\begin{aligned}
\frac{d\rho_{11}}{dt}&=-\Omega\rho^{I}_{12}+\Gamma_{21}\rho_{22}+\Gamma_{51}\rho_{55}+\Gamma_{61}\rho_{66}\\
\frac{d\rho_{22}}{dt}&=\Omega\rho^{I}_{12}-(\Gamma_{21}+\Gamma_{23}+\Gamma_{24})\rho_{22}\\
\frac{d\rho_{33}}{dt}&=\Gamma_{23}\rho_{22}-(\Gamma_{35}+\Gamma_{36})\rho_{33}\\
\frac{d\rho_{44}}{dt}&=\Gamma_{24}\rho_{22}-(\Gamma_{45}+\Gamma_{46})\rho_{44}\\
\frac{d\rho_{55}}{dt}&=\Gamma_{35}\rho_{33}+\Gamma_{45}\rho_{44}-\Gamma_{51}\rho_{55}\\
\frac{d\rho_{66}}{dt}&=\Gamma_{36}\rho_{33}+\Gamma_{46}\rho_{44}-\Gamma_{61}\rho_{66}\\
\frac{d\rho^{I}_{12}}{dt}&=\frac{1}{2}\Omega(\rho_{11}-\rho_{22})+\rho^{R}_{12}\Delta-\frac{1}{2}\Gamma_{21}\rho^{I}_{12}\\
\frac{d\rho^{R}_{12}}{dt}&=-\rho^{I}_{12}\Delta-\frac{1}{2}\Gamma_{21}\rho^{R}_{12}
\end{aligned}
\end{equation}

The subscript numbers of the density matrix correspond to different
energy levels shown in Fig.~1. The diagonal elements $\rho_{11}$,
$\rho_{22}$ and so on mean the probability of atoms in corresponding
states. $\rho^{R}_{12}$ means energy shift and $\rho^{I}_{12}$ means
power broadening. $\Delta=\omega_{21}-\omega$ is frequency detuning
of pumping light on the transition frequency and can be set to 0.
$\Gamma_{21}$, $\Gamma_{23}$ \emph{etc.} are decaying rates related
to the lifetimes of corresponding energy levels described in
Fig.~1.

The numerical solutions of the above equations are shown in Fig.~2.
It is obvious that under the action of pumping laser, the atoms at
$5S_{1/2}$ state decrease rapidly. The oscillation of $\rho_{11}$
and $\rho_{22}$ is caused by the assumption that the 421.7 nm
pumping laser is monochromatic. At steady-state, the value of
$\rho_{33}$ is 8.4\% and $\rho_{55}$ is 3.3\%. This means the atoms
in $6S_{1/2}$ state are about 2.5 times as many as that in
$5P_{3/2}$ state, therefore the population inversion is built up.
From Fig.~2 we can also conclude that the population inversion
between $6S_{1/2}$ and $5P_{3/2}$ states is built up at a time scale
of $10^{-6}$ s. Other population inversions between
$6S_{1/2}-5P_{1/2}$, $4D_{3/2}-5P_{3/2}$ and $4D_{3/2}-5P_{1/2}$
states are established simultaneously. We can experimentally detect
the fluorescence spectrum and calculate relative lines intensity to
demonstrate whether these population inversions are built up.

\section{LASING OF RUBIDIUM FOUR-LEVEL QUANTUM SYSTEM ACTIVE OPTICAL CLOCK}

An optical resonant cavity could be applied to detect the
probability of lasing transition between the inverted states
$6S_{1/2}$ and $5P_{3/2}$ as the population inversion occurs.
According to the classical laser theory, the oscillating process
could start up once the optical gain exceeding the loss rate. Here
we use a bad cavity ($\Gamma_{c}\gg\Gamma_{gain}$) to realize laser
output. When the stimulated radiation starts and reaches the
condition of self-oscillation, the laser oscillation is built up and then
output the 1366.9 nm laser of active optical clock which will be
used as laser frequency standards.

The atom-cavity coupling constant $g$~\cite{An1, An2} is given as
$g=\sqrt{\mu^{2}\omega_{35}/(2\hbar\varepsilon_{0}V_{mode})}$, where
$\omega_{35}=2\pi c/\lambda_{35}$ and $\mu^{2}=3\pi\hbar
c^{3}\varepsilon_{0}\Gamma_{35}/\omega_{35}^{3}$. Then the
atom-cavity coupling constant $g$ can be written as
$g^{2}=3c\Gamma_{35}\lambda_{35}^{2}/(8\pi V_{mode})$. The variation
range of $g$ is about from $3\times10^{4}$ s$^{-1}$ to
$9\times10^{5}$ s$^{-1}$ for reasonable value of $V_{mode}$. The
laser emission coefficient is
$\sin^{2}(\sqrt{n+1}g\tau_{int})$~\cite{An1, An2}, where
$\tau_{int}=1/(\Gamma_{3}+\Gamma_{5})=17.4$ ns considering the
condition of $\sqrt{n+1}g\tau_{int}\geq\pi/2$ in our case discussed
here. Define the cycle time for Rb atoms through the passage of
$5S_{1/2}$, $6P_{1/2}$, $6S_{1/2}$ and $5P_{3/2}$ states with the
action of 421.7 nm pumping laser during transition time as
\begin{equation}
\tau_{cyc}=\frac{1}{\Gamma_{2}}+\frac{1}{\Gamma_{3}}+\frac{1}{\Gamma_{5}}+\frac{1}{\Omega}
\end{equation}
and $\tau_{cyc}=225$ ns for $\Omega=2.64\times10^{7}$ s$^{-1}$
neglecting other channel. Thus the inverted atoms that are possible
to emit photons per unit time is
$(\rho_{33}^{'}-\rho_{55}^{'})/\tau_{cyc}$, where $\rho_{33}^{'}$
and $\rho_{55}^{'}$ represent the number of atoms in $6S_{1/2}$
state and $5P_{3/2}$ state. So the gain term of the rate equation
for cavity photon number $n$ is
$\sin^{2}(\sqrt{n+1}g\tau_{int})(\rho_{33}^{'}-\rho_{55}^{'})/\tau_{cyc}$.

The equations for emitted photons ($n$) from the total atoms inside
the cavity mode ($N$), together with corresponding density matrix
equations are listed as follows. Here we apply the form of
semiclassical approximation of four-level quantum system~\cite{An1,
An2}, then
\begin{equation}
\begin{aligned}
\frac{d\rho_{11}^{'}}{dt}&=-\Omega\rho^{I}_{12}+\Gamma_{21}\rho_{22}^{'}+\Gamma_{51}\rho_{55}^{'}+\Gamma_{61}\rho_{66}^{'}\\
\frac{d\rho_{22}^{'}}{dt}&=\Omega\rho^{I}_{12}-(\Gamma_{21}+\Gamma_{23}+\Gamma_{24})\rho_{22}^{'}\\
\frac{d\rho_{33}^{'}}{dt}&=\Gamma_{23}\rho_{22}^{'}-(\Gamma_{35}+\Gamma_{36})\rho_{33}^{'}-\frac{\rho_{33}^{'}-\rho_{55}^{'}}{\tau_{cyc}}\sin^{2}(\sqrt{n+1}g\tau_{int})\\
\frac{d\rho_{44}^{'}}{dt}&=\Gamma_{24}\rho_{22}^{'}-(\Gamma_{45}+\Gamma_{46})\rho_{44}^{'}\\
\frac{d\rho_{55}^{'}}{dt}&=\Gamma_{35}\rho_{33}^{'}+\Gamma_{45}\rho_{44}^{'}-\Gamma_{51}\rho_{55}^{'}+\frac{\rho_{33}^{'}-\rho_{55}^{'}}{\tau_{cyc}}\sin^{2}(\sqrt{n+1}g\tau_{int})\\
\frac{d\rho_{66}^{'}}{dt}&=\Gamma_{36}\rho_{33}^{'}+\Gamma_{46}\rho_{44}^{'}-\Gamma_{61}\rho_{66}^{'}\\
\frac{d\rho^{I}_{12}}{dt}&=\frac{1}{2}\Omega(\rho_{11}^{'}-\rho_{22}^{'})+\rho^{R}_{12}\Delta-\frac{1}{2}\Gamma_{21}\rho^{I}_{12}\\
\frac{d\rho^{R}_{12}}{dt}&=-\rho^{I}_{12}\Delta-\frac{1}{2}\Gamma_{21}\rho^{R}_{12}\\
\frac{dn}{dt}&=\frac{\rho_{33}^{'}-\rho_{55}^{'}}{\tau_{cyc}}\sin^{2}(\sqrt{n+1}g\tau_{int})-\Gamma_{c}n
\end{aligned}
\end{equation}

One should note that the variables $\rho_{11}^{'}$ \emph{etc.} in
these equations, which represent the number of atoms in the
corresponding states, are different from those in equation (1). At
room temperature the number of Rb atoms in the cavity mode is about
$N=7\times10^9$ with $V_{mode}=10^{-7}$ m$^{3}$. $N$ can be changed
through the adjustment of temperature and $V_{mode}$. At
$120\,^{\circ}\mathrm{C}$, there are about $2\times10^{12}$ atoms
inside the cavity mode with the same mode volume. $\rho_{11}^{'}=N$
and others are all 0 at the beginning. The cavity loss rate
$\Gamma_{c}=a\Gamma_{gain}$, where $\Gamma_{gain}$ =$\Gamma_{sp}$ is
the natural linewidth of clock transition for laser cooled atoms in
MOT, and $\Gamma_{sp}$ equals to 12.9 MHz here. For atoms in thermal
gas cell, we assume the coherent 421.7 nm pumping laser only pumps
atoms with velocity near zero for simplicity, then the
$\Gamma_{gain}$ $\simeq$$\Gamma_{sp}$ holds approximately. To reduce
the effect caused by the thermal noise of cavity, the cavity loss
rate $\Gamma_{c}$ should be far greater than the natural linewidth
of output laser ($\Gamma_{35}$). We set $a=100$ in this paper.

\begin{figure}[!h]
\label{fig:Figer3}
\includegraphics[width=7.5cm]{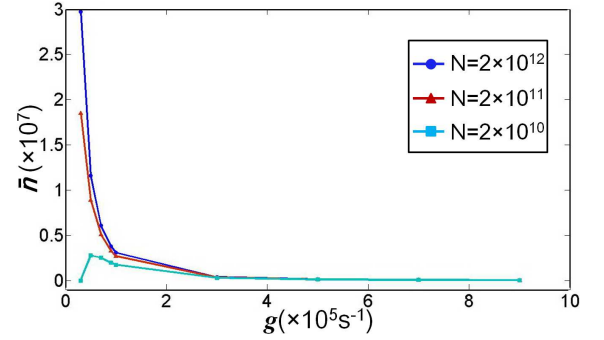}
\caption{(Color online) The photon number at steady-state inside the cavity varies with the atom-cavity coupling constant $g$ under different value of $N$.}
\end{figure}

\begin{figure}[!h]
\label{fig:Figer4}
\includegraphics[width=7.5cm]{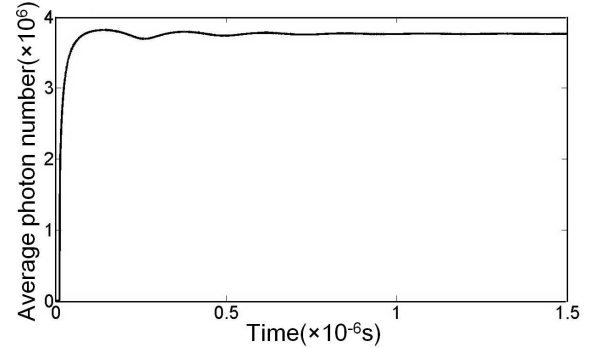}
\caption{(Color online) The average photon number inside the cavity under the condition of
$\Omega=2.64\times10^{7}$ s$^{-1}$, $g=9\times10^{4}$ s$^{-1}$ and $N=2\times10^{12}$.}
\end{figure}

Fig.~3 describes the solutions for the photon number equation. The
steady-state value of photon number $\overline{n}$ decrease with the
enhancement of atom-cavity coupling constant. The increase of total
atoms inside the cavity mode has less influence over $\overline{n}$ compared
to the variation of $g$. Fig.~4 describes the dynamical lasing
process with the action of pumping laser. We can conclude that a
stable laser field is built up within laser cavity on condition that
the population inversion is preserved. The steady-state photon
number $\overline{n}=3.76\times10^{6}$ for $g=9\times10^{4}$
s$^{-1}$ ($V_{mode}=10^{-7}$ m$^{3}$) and $N=2\times10^{12}$. The
corresponding output laser power $P=\overline{n}h\nu_{35}\Gamma_{c}$
is 0.7 mW.

Since $\Gamma_{3},
\Gamma_{5},\Gamma_{35}<\tau_{int}^{-1}\ll\Gamma_{c}$, the quantum
linewidth of the active optical clock can be approximated to be
$D=g^{2}/\Gamma_{c}$~\cite{Yu}. Therefor we can get the
quantum-limited linewidth of Rb four-level quantum system active
optical clock in the magnitude of Hz level. If $g=3\times10^{4}$
s$^{-1}$ and $\Gamma_{c}=1.29\times10^{9}$ s$^{-1}$, the linewidth
is 0.7 Hz.

In consideration of the fact that the 421.7 nm pumping laser is far
detuning from the 1366.9 nm clock transition laser, the light shift
of $6S_{1/2}-5P_{3/2}$ clock transition caused by the 421.7 nm
pumping laser can be written as $\Omega^{2}/\Delta^{'}$, where
$\Delta^{'}$ is the detuning which equals $3.77\times10^{15}$ Hz.
Given the Rabi frequency $\Omega=2.64\times10^{7}$ s$^{-1}$, the
light shift will be 0.18 Hz. The stability of 421.7 nm pumping laser
power can be better than $10^{-3}$, therefore the uncertainty of
light shift due to pumping laser is less than 0.18 mHz. This means
to four-level quantum system, the influence on system stability and
accuracy induced by light shift is far less than that of two-level
and three-level quantum systems. Considering the mature technology of laser
cooling and trapping of Rb atoms, one can establish a 421.7 nm Rb
MOT directly for Rb active optical clock, and the cooling and
trapping laser will play the pumping role also since its light shift
is small enough.

\section{EXPERIMENTAL DESIGN OF ACTIVE OPTICAL FREQUENCY STANDARDS BASED ON RUBIDIUM FOUR-LEVEL
QUANTUM SYSTEM}

The lasing of Rb four-level quantum system active optical clock can be realized by using a
conventional Fabry-Perot resonator. But the difference is that the cavity loss rate $\Gamma_{c}$ far
outweighs the natural linewidth $\Gamma_{sp}$.

\begin{figure}[!h]
\label{fig:Figer5}
\includegraphics[width=7.5cm]{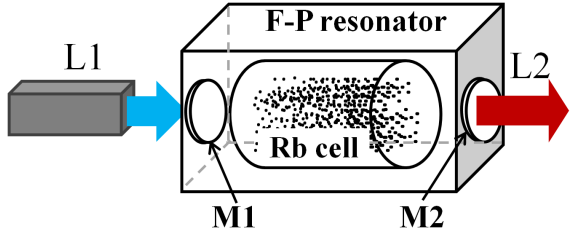}
\caption{(Color online) The experimental configuration of Rb four-level quantum system active optical clock.
L1 is the 421.7 nm pumping laser source. L2 is the 1366.9 nm output clock transition laser. M1 and M2 are coated mirrors.}
\end{figure}

The length of this cavity $L$ is set to 1.5 cm, then the free spectral range is
$FSR=c/(2L)=10$  GHz. The fineness
\begin{equation}
F=\frac{FSR}{\Gamma_{c}}=\frac{\pi\sqrt{r_{1}r_{2}}}{1-r_{1}r_{2}}=\frac{\pi\sqrt{R}}{1-R},
\end{equation}
where $R=r_{1}r_{2}$. Given $\Gamma_{c}=1.29$ GHz, from above
equation we can get $F=7.75$. Thus we can determine $R=0.67$,
$r_{1}=1$ and $r_{2}=0.67$. M1 mirror is coated with 421.7 nm
anti-reflection coating and 1366.9 nm high-reflection coating. M2
mirror is coated with 421.7 nm high-reflection coating and the
reflectivity of 1366.9 nm is 67\%. The final output laser comes out
from M2. Compared with other optical clock schemes, there are two
interesting points of this experimental scheme. The first one is
that the most parameters of atomic structure of alkali metals
including Rb have been measured precisely already. The second one is
that the laser cooling and trapping technology of alkali metals
including Rb is currently very mature, thus the thermal gas cell can
be replaced by magneto-optical trap for reducing the influence of
Doppler effect. The center frequency of four-level quantum system
active optical clock is decided by the transition frequency of Rb
atoms, but not the laser cavity mode which is very sensitive to the
vibration and fluctuation of temperature. The cavity mode of active
optical clock can be locked to a reference cavity with low thermal
noise~\cite{Kessler}, then the cavity pulling reduced by a factor
of $a=\Gamma_{c}/\Gamma_{gain}$ will be smaller than mHz.

\section{DISCUSSION AND CONCLUSION}
Active optical clock, whose quantum-limited linewidth of output
laser is far narrower than the natural linewidth of atomic spectrum
and its center frequency is not directly subject to cavity mode
noise, can provide optical frequency standards with high stability
and accuracy. Considering the accuracy requirement, lasing of a
three-level conventional laser is not so suitable for an active
optical frequency standards because the light shift due to the high
intensity pumping laser can't be tolerable. Thus it needs special
design of pumping scheme for active optical clock with three-level
configuration~\cite{Chen2, Meiser1} .

Four-level quantum system active optical frequency standards not only possesses the advantages of active clock such as
narrow linewidth and high stability and accuracy, but also overcomes the limitation on stability and accuracy due to light shift.
Rb atom four-level quantum system, as a typical example, is studied in detail.
Other alkali metals including potassium and cesium have similar four-level quantum systems structures and therefore are appropriate candidates for this novel scheme
of active optical clock. The parameters of atomic structure of these atoms have been measured precisely already thus provides great convenience.
Take cesium as an example, the levels $6S_{1/2}$, $7P_{3/2}$, $7S_{1/2}$
and $6P_{3/2}$ compose a four-level quantum system suitable for active optical frequency standards.
Population inversion in cesium atoms has been achieved in our preliminary experiment.
It is expected to realize active optical clock based on four-level quantum system in our next experimental study.

We mainly study atoms in thermal gas cell in this paper. It is
valuable to notice that four-level quantum system based on atoms in
MOT can reduce the influence of Doppler effect greatly. To reduce
the effect of collision shift and decrease Doppler effect further,
magic wavelength lattice trapped atoms can be applied. We have
identified several magic wavelengths for Rb four-level quantum
system active optical clock~\cite{Zang}. Besides, laser cooling and
trapping technology of alkali metals is very mature. So active
optical lattice clock based on four-level quantum system atoms is
expected to greatly improve the properties of atomic clocks.

In summary, we propose a scheme of four-level quantum system active
optical clock with Rb atoms. Our
calculations show that the population inversion between $6S_{1/2}$
and $5P_{3/2}$ states of Rb atom can be built up at a time scale of
$10^{-6}$s. With 421.7 nm pumping laser operating at $5S_{1/2}$ to
$6P_{1/2}$ transition and $2\times10^{12}$ atoms inside the cavity
mode, the output laser power of Rb four-level quantum system active
optical clock can reach 0.7 mW for $\Gamma_{c}=100\Gamma_{35}=1.29$
GHz. The quantum-limited linewidth can reach 0.7 Hz. The stability
of light shift can be smaller than 0.18 mHz with the Rabi frequency
of pumping laser $2.64\times10^{7}$ s$^{-1}$. Other alkali metals
like potassium and cesium are also suitable for four-level quantum
system active optical clock. An experimental configuration to
realize active optical frequency standards based on Rb four-level
quantum system is also given.

\section{ACKNOWLEDGMENT}
The authors thank Shengnan Zhang, Zhichao Xu and Zhong Zhuang for helpful discussions.
This work is supported by the National Natural Science Foundation of China (grants 10874009 and 11074011).

\end{document}